\documentclass[11pt]{article}
\usepackage[utf8]{inputenc}
\usepackage[margin = 1.25in]{geometry}
\usepackage{graphicx, setspace, subcaption}
\usepackage{amsfonts, amsmath, amssymb, amsthm}
\title{What Motivated Mitigation Policies? A Network-Based Longitudinal Analysis of State-Level Mitigation Strategies}
\author{William Fries\\\small{GIDP in Applied Math}\\\small{University of Arizona}}
\date{December 2022}

\begin{document}

\maketitle
\begin{abstract}
    Understanding which factors informed pandemic response can help create a more nuanced perspective on how each state of the United States handled the crisis. To this end, we create various networks linking states together based on their similarity in mitigation policies, politics, geographic proximity and COVID-19 case data. We use these networks to analyze the correlation between pandemic policies and politics, location and case-load from January 2020 through March 2022. We show that the best predictors of a state's response is an aggregate political affiliation rather than solely governor affiliation as others have shown \cite{governor_outcomes, governor_and_cases}. Further, we illustrate that political affiliation is heavily correlated with policy intensity from June 2020 through June 2021, but has little impact on policy after June 2021. In contrast, geographic proximity and daily incidence are not consistently correlated with state's having similar mitigation policies.
\end{abstract}
\section{Introduction}

Throughout the COVID-19 pandemic, many in the United States turned to their state leaders for guidance and to instate policies that best served their needs. In the midst of a global crisis, we should expect that pandemic intensity dictates the policies that political leaders implement to protect their citizens, whether medically, economically or otherwise. However, throughout the course of the pandemic, it became evident that politics played a much more important role than the severity of the crisis itself. Many have performed statistical analyses that illustrate the strong correlation between political affiliation, the policies instated during the pandemic \cite{governor_outcomes, governor_and_cases, adeel2020covid}, and the subsequent outcomes \cite{polarization,hsiang2020effect,flaxman2020estimating}. Others have examined how the citizens have reacted to these policies \cite{Franz2021,defranza_lindow_harrison_mishra_mishra_2021,gadarian_goodman_pepinsky_2021, painter2020political} or vaccine uptake rates based on a number of factors, including political affiliation \cite{druckman2021role, levin2022faith, levin2022determinants, bolsen2022politicization}

However, few of these studies analyzed to what extent the relationships between political actors plays a role. \cite{governor_outcomes} and \cite{governor_and_cases} perform longitudinal studies that analyze the impact of governor policy through November and December 2020 respectively. This analysis does not span the whole pandemic and these results only consider the governor's political affiliation. While this intuitively makes sense, as governors implemented executive orders, states such as Maryland and Massachusetts, which are more liberal, are classified as Republican states. This is recognized as a limitation of the study and we hope to quantify how much of a limitation this could be and offer a remedy. 

We hope to further expand on \cite{governor_outcomes} and \cite{governor_and_cases} by considering when the political influence is statistically significant and under what conditions politics plays a strong role. Studies have shown that both homophily and proximity can be important factors in political networks and their resulting outcomes \cite{gerber_henry_lubell_2013, pietryka_debats_2017}. We analyze to what extent homophily could play a role in the mitigation policy process. That is, do Republican and Democrat states tend to cluster together in their approach to handling the pandemic or is there still variation. In contrast, we consider to what effect geographic proximity and current pandemic intensity can be used to explain each state's mitigation response. As both of these factors, specifically the latter, would be expected to have a strong impact on mitigation policy strength, we might expect to see a high correlation. As states with large amounts of new cases would likely intensify mitigation policies; similarly, as the pandemic begins to weaken, so might the policies. This acts as our expectation in the analysis below. 

Through analyzing the pandemic responses through a network-based lens, we seek to understand how the political similarity, geographic proximity and pandemic intensity between states might correlate to the overall mitigation policies throughout the pandemic and when they played a significant role.

\section{Methods}
To best analyze how these factors impacted the mitigation strategies implemented by each state we first quantify each states general mitigation policy strength. To quantify the mitigation policy strength, we use separate mitigation strategies as defined in \cite{Google,Oxford} for dates ranging from Jan. 22, 2020 to March 28, 2022. We consider the following mitigation strategies in our analysis: school closing, workplace closing, cancelling of public events, restrictions on gatherings, public transport closing, stay at home requirements, restrictions on internal movement, debt relief, and facial coverings.

Each of these measures is rated on a scale to indicate the severity. Table \ref{tab:policy} describes each strategy and details can be found in \cite{Google}. For consistency we normalize the value in column 2 to range from 0 to 1 where 0 indicates no measure and 1 indicates the highest possible intensity. These measures will be used for both standard statistical analysis as well as in constructing networks to potentially uncover more nuanced perspectives. 

To quantify political affiliation, we use four separate measures: governor affiliation, state senate majority, state legislative majority and 2020 electoral college. We will use these to analyze which political markers best predict mitigation policy intensity and construct a Political Adjacency Network.

\begin{table}[]
    \centering
    \begin{tabular}{l|l|l}
        Name & Range &	Description	\\ \hline
        school closing	& [0-3] &	Level to which schools are closed \\
        workplace closing	& [0-3]	& Level to which workplaces are closed 	\\
        cancel public events	& [0-3] &	Amount of restriction on public events\\
        restrictions on gatherings	& [0-3] & Restricted gatherings of non-household members 	\\
        public transport closing	& [0-3]	& Amount of public transportation that is closed	\\
        stay at home requirements	& [0-3]	& Who self-quarantine at home is mandated for\\
        restrictions on internal movement	& [0-3] & Amount of travel restriction within country \\
        debt relief	& [0-3]	& Amount of debt/contract relief for households	\\
        facial coverings	& [0-4] &	Policies on facial coverings outside the home	\\
    \end{tabular}
    \caption{Descriptions of mitigation policies used in the quantifying the mitigation strength of each state. Detailed explanations for each measurement criteria can be found at \cite{Oxford}. The data was downloaded from the Google open-source COVID-19 data repository at \cite{Google}.}
    \label{tab:policy}
\end{table}

\subsection{Network Construction}\label{sec:networks}
Network analysis offers a method to analyze the underlying relationships that might motivate individual agents to act \cite{Newman2010}. To best analyze which factors might impact mitigation policies on a relational level, we will consider how graphs representing these ideas are correlated with each other. As we seek to uncover potential correlations with mitigation strategy, we consider three separate variables: political affiliation, geographic proximity, and COVID-19 incidence rates. 

To generate the political network, we can define $P(s)$ to be the political strength of each state. For state $s_i$, 
\begin{equation}\label{eq:pol}
P(s_i) = p_{gov}(s_i) + p_{sen}(s_i) + p_{leg}(s_i) + p_{elec}(s_i)
\end{equation}
\begin{equation}
p_j(s_i) = \begin{cases}
1 & \text{if Democrat} \\
-1 & \text{if Republican} \\
2k-1  & \text{else}
\end{cases}
\end{equation}

where $k$ is the proportion of the political strength that is Democrat and where $p_j$ refers to the political body measured (eg. governor affiliation). For example, if a state has a Republican governor, Republican dominated legislature, a 50/50 split senate and voted Republican in the 2020 election then 
\begin{equation}
    P(s_i) = -1 + 0 - 1 - 1.
\end{equation}
We then connect states $s_i$ and $s_j$ if $\|P(s_i) - P(s_j)\| \leq 2$. 

To illustrate when connections are made, Missouri is a fully Republican state, then $P(\text{MO}) = -4$. North Carolina is Republican dominated but has a Democrat governor, thus $P(\text{NC}) = -2$. Finally, Pennsylvania is split with a Republican congress but a Democrat governor and electoral college. Thus $P(\text{PA}) = 0$. So, North Carolina will be connected to Missouri and Pennsylvania; however, Missouri will not be connected to Pennsylvania.

We can define an incidence network to illustrate states which have comparable increase in COVID-19 cases. We define $\mathcal{I}_{s}(t)$ to be the per-capita incidence of state $s$ at time $t$. We then draw an edge between state $s_i$ and $s_j$ if 

\begin{equation}
\|\mathcal{I}_{s_i}(t)-\mathcal{I}_{s_j}(t)\| < \frac{0.05}{50} \sum_{k=0}^{49}\mathcal{I}_{s_k}(t) = 0.001\sum_{k=0}^{49}\mathcal{I}_{s_k}(t)
\end{equation}

We consider a new network for every time step for both the mitigation similarity network and the incidence similarity network.

Finally, we create a geographic network of the 50 states. In this network states are connected if they share a border. This generates a time-independent network with three components (the contiguous United States, Alaska, and Hawaii).

To collectively analyze the mitigation strategies for each state, we consider a generalized mitigation policy strength. Let $m_s^i(t)$ be the mitigation strategy $i$ at time $t$ for state $s$. Then the mitigation strength for state $s$ at time $t$ will be defined as $M_s(t) := \|\mathbf{m_s(t)}\|_2$ where $\mathbf{m_s(t)} = \langle m_s^0(t), \dots m_s^8(t)\rangle$ and $\|\cdot\|_2$ denotes the Euclidean norm. Finally, we define $\boldsymbol\mu(t)$ to be the mean mitigation policy strength at time $t$ across all 50 states. Thus, the $i$th component of $\boldsymbol\mu(t)$ is 

\begin{equation}
    \mu_t^i = \frac{1}{50}\sum_{s=0}^{49}m_s^i(t).
\end{equation}

Using $\boldsymbol\mu(t)$, we define a network of states in which an edge represents comparable strength of mitigation policy. An edge exists between states $s_i$ and $s_j$ if 

\begin{equation}
    \|\mathbf{m_{s_i}(t)} - \mathbf{m_{s_j}(t)}\|_2 < 0.05\|\boldsymbol\mu(t)\|_2.
\end{equation}

\begin{figure}
    \centering
    \begin{subfigure}[b]{.45\linewidth}        
        \includegraphics[width=\textwidth]{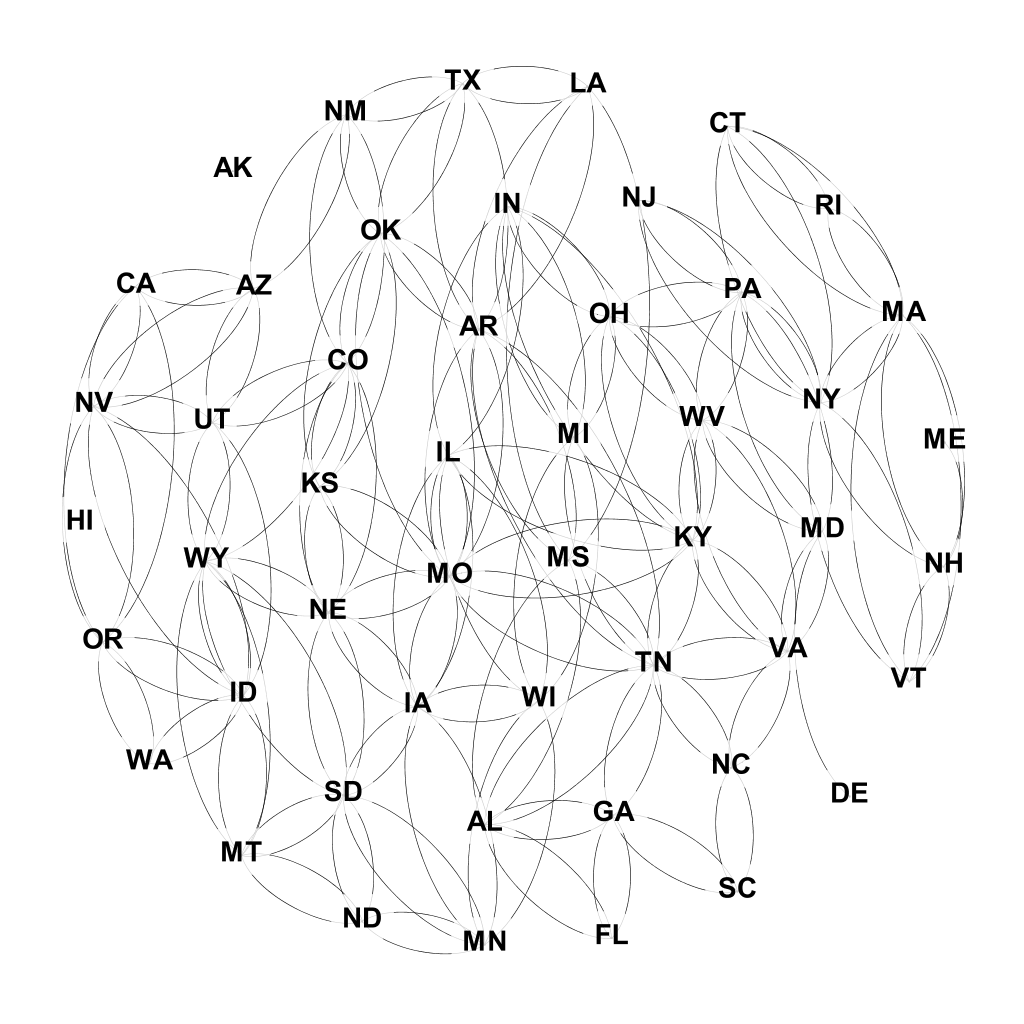}
        \caption{Geographic Adjacency Network}
        \label{fig:geo}
    \end{subfigure}
    \begin{subfigure}[b]{.45\linewidth}
        \includegraphics[width=\textwidth]{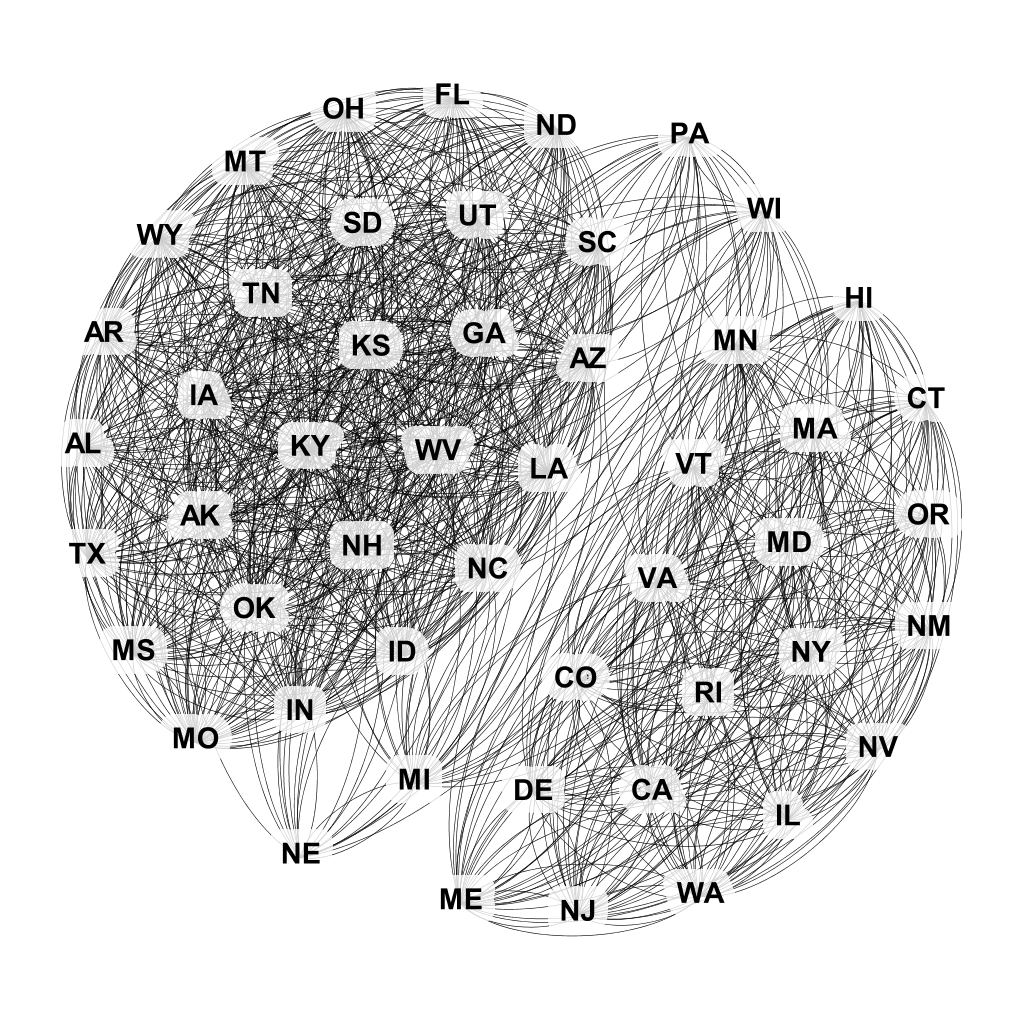}
        \caption{Political Adjacency Network}
        \label{fig:poli}
    \end{subfigure}
    \begin{subfigure}[b]{.45\linewidth}
        \includegraphics[width=\textwidth]{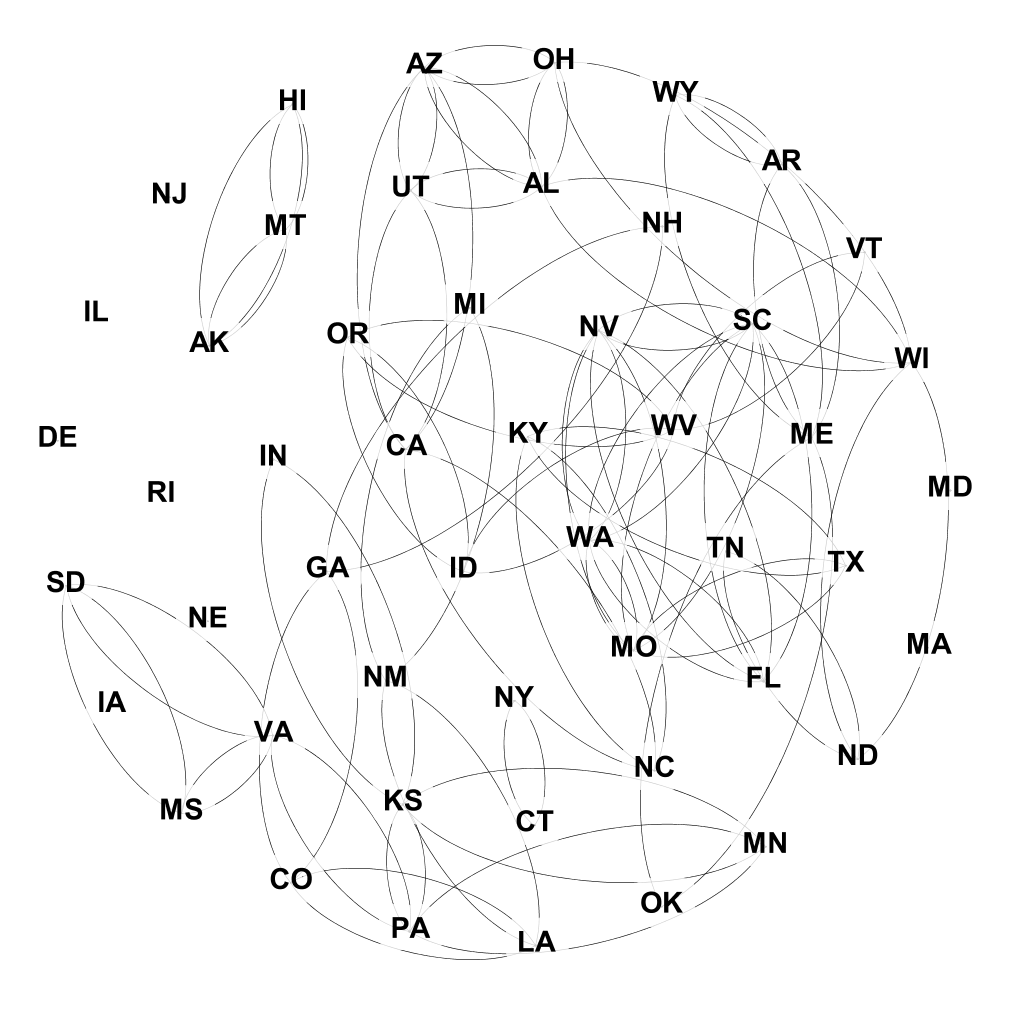}
        \caption{Incidence Similarity Network}
        \label{fig:inc}
    \end{subfigure}
    \begin{subfigure}[b]{.45\linewidth}
        \includegraphics[width=\textwidth]{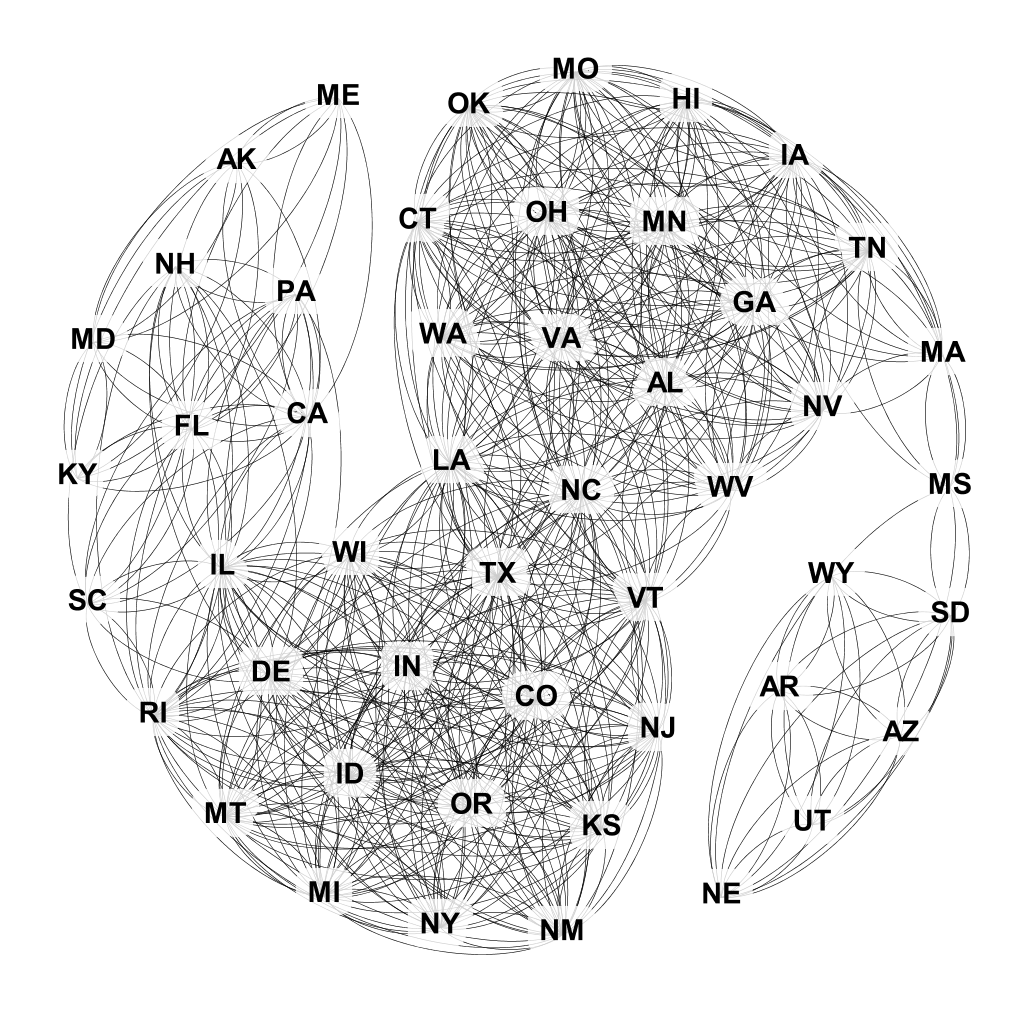}
        \caption{Mitigation Similarity Network}
        \label{fig:mit}
    \end{subfigure}
    \caption{Examples of the four networks used in the analysis. Both the Incidence Similarity and Mitigation Similarity networks are taken on May 1, 2020. Visualizations were created using Gephi \cite{gephi}.}
    \label{fig:network examples}
\end{figure}

In our network-based analysis, we will use these two static networks (geographic and political) and dynamic network (incidence similarity) to investigate the correlation with the mitigation similarity networks. An example of each of the networks can be found in Figure \ref{fig:network examples} where, for the two dynamic networks, we take the network from May 1, 2020.

\subsection{Statistical Methods}

We then consider multiple methods to quantify the correlation between these factors and overall COVID mitigation policy. As done in \cite{governor_outcomes, governor_and_cases}, we consider the correlation between political alignment and mitigation strength. At every time step, we compute this Pearson correlation coefficient with respect to each of the four individual political measures and the aggregate political measure. This allows us to analyze how the political influence might have evolved over the course of the pandemic. The results to this analysis can be found in Section \ref{sec:results correlation}. 

To determine potential underlying relationships between similar mitigation policies and the three explanatory variables, we considered a graphical model and quadratic assignment procedure (QAP) \cite{hubert1976quadratic, krackardt1987qap, krackhardt1988predicting, lisette-espin}. This method extends standard graph correlation by controlling for graph structure. Standard graph correlation examines the correlation between the edges of two networks. However, QAP analysis allows us to consider only graphs with the same structure, making the result dependent on the specific relationships rather than the underlying structure. In cases where one of the two graphs might be significantly sparser than the other, correlating the edges will likely not yield high values as the number of edges is significantly different. 

QAP remedies this problem. For every mitigation similarity network, we compute the correlation of the edges with those of the political, geographic and incidence similarity networks. We then permute the node (ie. state) labels, compute the new correlation, and repeat this for all possible permutations of node labels. This results in a distribution of correlations. This distribution represents the correlations between two networks for all isomorphic graphs of this structure. We can then ask the correlation of the original graph labels compares to those of isomorphic graphs. Using the corresponding p-value, we judge the significance of the correlation of the original graph. We use this method to compare the political similarity, geographic proximity, and incidence similarity networks with the mitigation similarity networks.

\section{Results}
We explore the correlation between political affiliation and mitigation policy intensity and the correlation between the various networks described in Section \ref{sec:networks}. We split each of these analyses into separate sections and discuss their combined implications in Section \ref{sec:discussion}.

\subsection{Pearson Correlation Coefficient}\label{sec:results correlation}
\begin{figure}
    \centering
    \includegraphics[width = \linewidth]{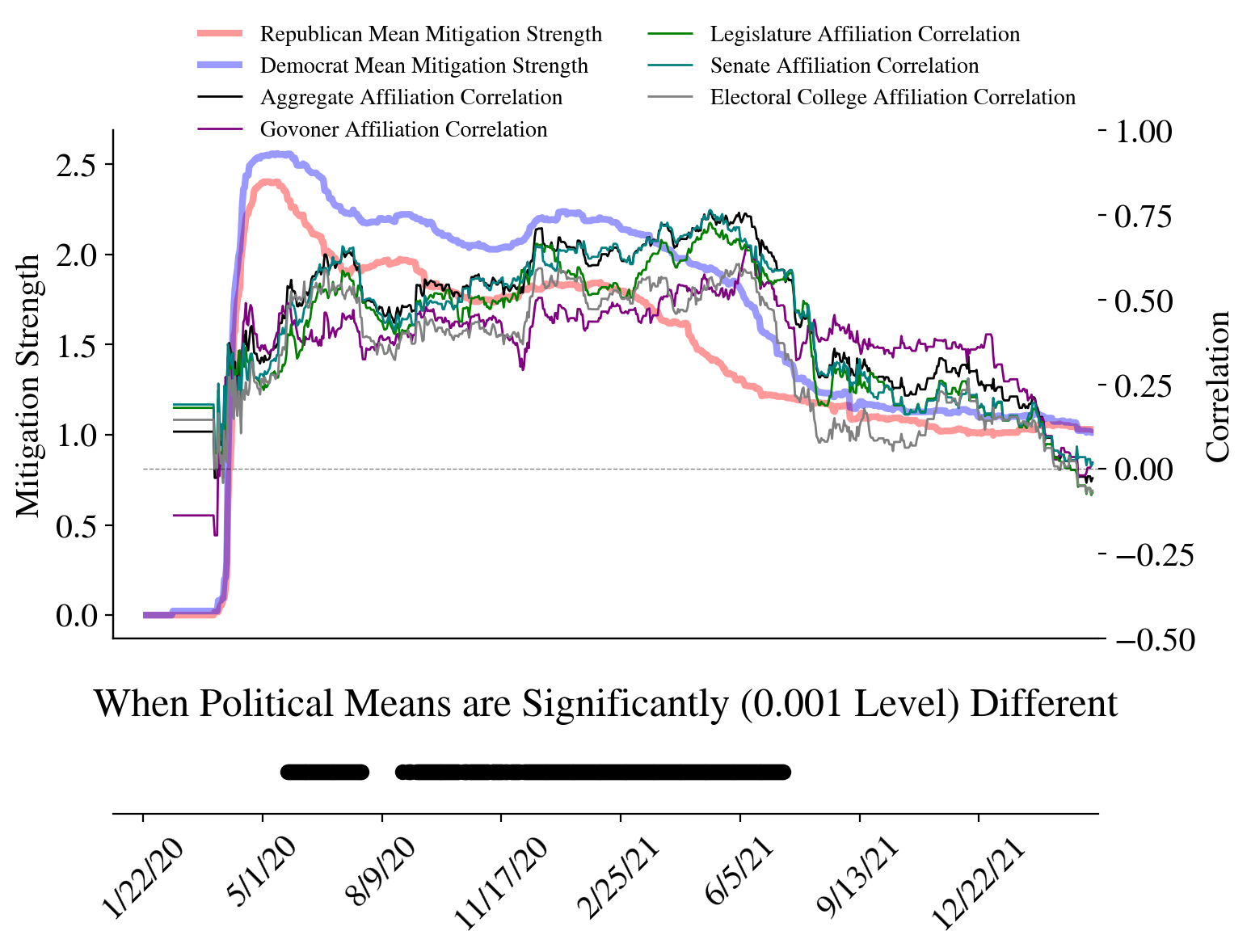}
    \caption{Correlation of Mitigation policy strength and political affiliation measured by governor affiliation, senate and legislative majority, and 2020 electoral college and when there is statistically significant difference in mitigation policy intensity between majority Democratic and Republican states.}
    \label{fig:correlation}
\end{figure}

For each day of data, we track the correlation between mitigation policy strength and the four political measures and the aggregate measure as defined in Equation \ref{eq:pol}. The results are shown in Figure \ref{fig:correlation}. 

Through the course of the pandemic, we see there is a general positive correlation between mitigation strength and all measures of state-level political affiliation. We report the average values of this correlation over two separate time periods in Table \ref{tab:pearson_corr}. Interestingly, throughout most of the first year of the pandemic, governor affiliation has the lowest correlation amongst all political measures, with an average correlation of 0.44 and a maximum of 0.58. As governors signed executive orders, we might expect this correlation to be the highest but the analysis illustrates otherwise. This contradicts the idea that governors affiliation would be the best predictor. However, as mitigation strategies relaxed in mid-2021, governor affiliation become the best predictor of mitigation strength.

Further, the best predictor over the first year of the pandemic is either the overall party strength or the senate majority. On average, from January 2020 through March 2022, the overall political affiliation of the state was the best predictor of mitigation policy strength. We will discuss potential reasons for this in Section \ref{sec:discussion}. 

\begin{table}[]
    \centering
    \begin{tabular}{|c|c|c|c|}
    \hline
    \textbf{Affiliation} & \begin{tabular}{c}
         \textbf{Max}\\
         \textbf{Correlation}
    \end{tabular} & 
    \begin{tabular}{c}
         \textbf{Mean Correlation}\\
         \textbf{June 2020 - June 2021}
    \end{tabular}
    &
    \begin{tabular}{c}
         \textbf{Mean Correlation} \\
         \textbf{July 2021 - March 2022}
    \end{tabular}\\ \hline
        Aggregate & 0.76 & 0.57 & 0.24\\ \hline 
        Governor & 0.58 & 0.44 & 0.25 \\ \hline 
        Legislature & 0.72 & 0.51 & 0.19 \\ \hline
        Senate & 0.77 & 0.56 & 0.22 \\ \hline
        \begin{tabular}{c} Electoral \\ College \end{tabular}
               & 0.61 & 0.45 & 0.13 \\\hline
    \end{tabular}
    \caption{Correlation statistics for political affiliation and mitigation policy strength.}
    \label{tab:pearson_corr}
\end{table}

We perform a 2-sample t-test on the mean mitigation policy strength for both majority Republican and Democrat states. They are statistically significantly different during two different periods: May 22nd, 2020 through July 23rd, 2020 and from August 26th, 2020 through July 11th, 2021. This acts as an indicator that there is a meaningful relationship between political similarity and similar mitigation policies. We will discuss how this relates to the timing of significance of the network analysis in Section \ref{sec:discussion}.

\subsection{Network Correlation}

\begin{figure}
    \centering
    \includegraphics[width = \linewidth]{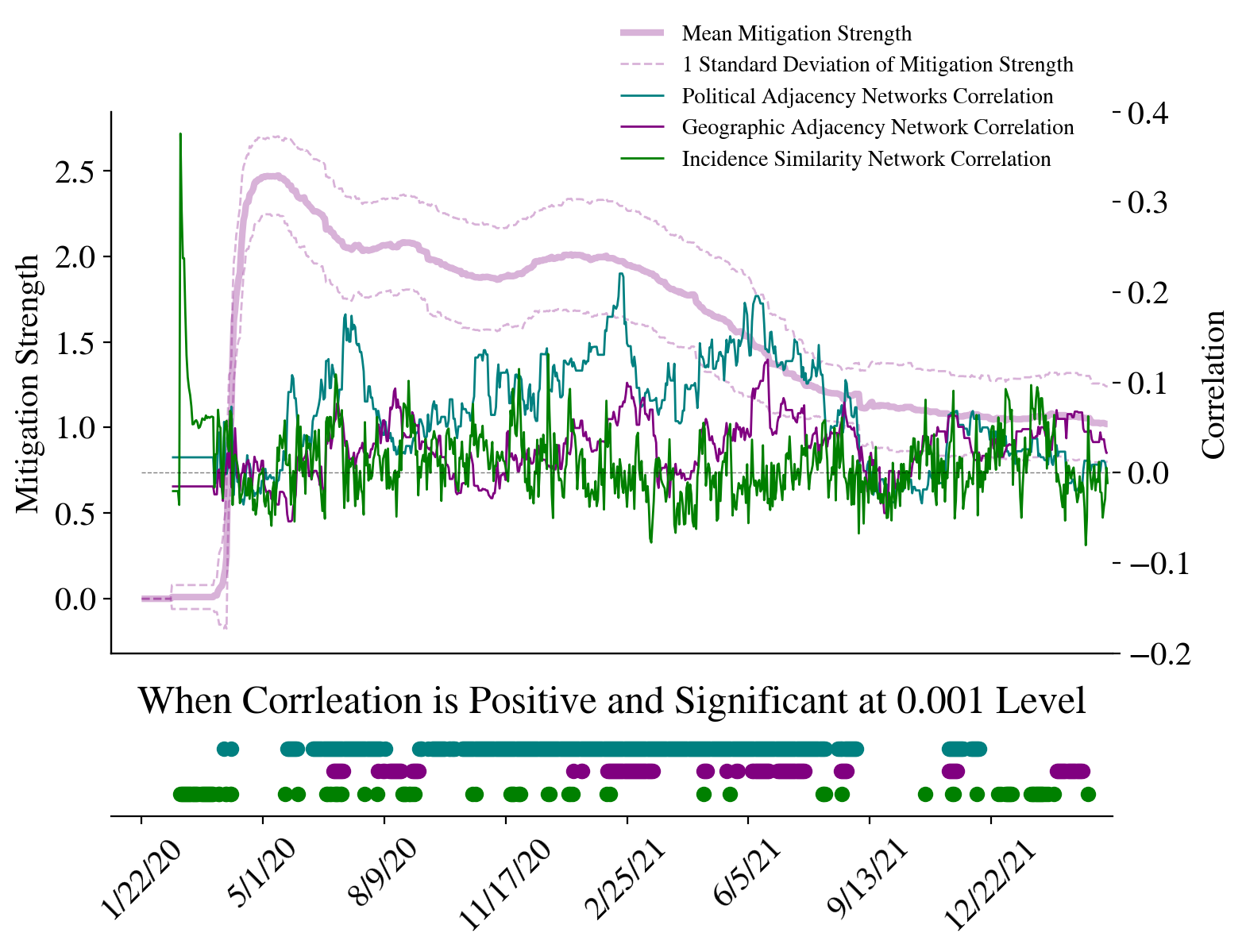}
    \caption{Correlation and regions of significance for QAP analysis for network representations of political similarity, geographic proximity and incidence similarity.}
    \label{fig:graph correlation}
\end{figure}

When considering the graph correlation through the QAP analysis, we look for a positive and significant correlation between the graphs. A positive and significant correlation would imply that the graph structures mirror each other and that one graph can be used to predict the structure of the other. As the number of edges between the graphs could vary greatly, we do not expect high correlation values. For example, the Political Adjacency Network contains 1152 edges whereas the average number of edges for the mitigation networks between June 2020 and June 2021 is 406 edges. In spite of differing numbers of edges, QAP allows us to consider when the correlation is significant, and thus, the network structures are significantly related. We set an $\alpha$-level of 0.001 and illustrate the results in Figure \ref{fig:graph correlation}.

Table \ref{tab:graph_corr} shows the data found in Figure \ref{fig:graph correlation}. We report the average correlation and number of days of significance between June 2020 and June 2021. We note that the period of significance for these networks extends beyond June 2021; however for consistency with Table \ref{tab:pearson_corr}, we report data between June 2020 and June 2021. We notice that the only graph that has significant correlation with the Mitigation Similarity Networks over a long period of time is the Political Adjacency Network. The Political Adjacency Network not only has the highest correlation between June 2020 and June 2021, but also the proportion of days which the political similarity aligns with the mitigation strength of a state.  Interestingly, the Geographic Adjacency Network appears to correlate with the Mitigation Similarity Networks more consistently than the Incidence Similarity Networks. Visually, this can be seen by the lack of regions of significance in at the bottom of Figure \ref{fig:graph correlation}. Concretely, the Geographic Adjacency Network was significantly correlated with the Mitigation Similarity Network for only 25\% of the days between June 2020 and June 2021. More striking, the Incidence Similarity Network is only significantly correlated with the Mitigation Similarity Network 10\% of the days during the same time period. This implies that neither geographic relationship nor the relative epidemic-severity of can be used to explain why states had comparably strong mitigation policies.
\begin{table}[]
    \centering
    \begin{tabular}{|c|c|c|}
    \hline
    \textbf{Network} &  
         \textbf{Mean Correlation}
    &
    \begin{tabular}{c}
         \textbf{Proportion of Days} \\
         \textbf{Significant Correlation}
    \end{tabular}\\ \hline
        Political Adjacency  & 0.10 & 0.87\\ \hline 
        Geographic Adjacency & 0.03 & 0.24 \\ \hline 
        Incidence Similarity & 0.01 & 0.10 \\ \hline
    \end{tabular}
    \caption{Graph correlation statistics between June 2020 and June 2021. We report the average graph correlation between the each of the explanatory graphs and the Mitigation Strength Adjacency Network during this time period and the number of days which this correlation is significant.}
    \label{tab:graph_corr}
\end{table}
\section{Discussion}\label{sec:discussion}

There are two interesting points of discussion from the analyses above: the distinctions between the methods and the implications of the results. The two methods have yielded the same results of many other analysis prior to this one: there is a significant relationship between a state's politics and how they responded to the COVID-19 pandemic. However, now, we know \textit{when} this relationship was significant. Both methods reported that the relationship was significant between May of 2020 and July of 2021. However, each method offered slightly different results. 

The strategy of analyzing Pearson Correlation of mitigation policy strength with political affiliation highlights a potential method for analyzing state-level politics more accurately. As the governor's affiliation showed the lowest correlation with general mitigation policy strength, it is important to have a more holistic view of a states politics. For instance, classifying states by governor during the pandemic would classify Maryland and Massachusetts as Republican states when every other aspect of their politics is decisively Democrat.

When considering the timing of the significance of the two-sample t-test, we see that the difference in responses between Republican and Democrat states became significant at the beginning of June. While we cannot say for certain what caused this shift, there are a few possibilities. Many states were ending their lock-downs and stay-at-home requirements. Political pressure from other politicians or constituents could have also caused this divergence of the political parties.

However, when considering the end of the period of significance in 2021, we can potentially attribute this to the general drop in mitigation policy strength. As the mean mitigation policy strength (purple in Figure \ref{fig:graph correlation}) shifts from 2 to near 1 over a 6 month period. This supports the hypothesis that higher restrictions and strong policies are more politically divisive. 

When comparing the two-sample t-test results to those of the political similarity network and mitigation similarity networks, we see that the network analysis gives a slightly broader region of significance. This is likely because it accounts for the relationships between states and their similar behaviors rather than looking at each state in isolation. This, in turn, emphasizes the importance of political relationships in the Mitigation Similarity Networks structures.

In contrast, the Geographic Adjacency Network and Incidence Similarity Network show little correlation to the Mitigation Similarity Networks. We notice that the Geographic Adjacency Network shows small regions of sustained significance, this is likely because of the political landscape of the United States: Democrat states tend to be on the coasts and Republican states cluster in more central states.

Interestingly, the Incidence Similarity Networks correlations are significant at the beginning of the pandemic, when no other network shows significant correlation. This is likely because all states were responding to the pandemic rather than their political counterparts. However, the significant correlations end at the end of April as many lockdown measures were beginning to be lifted.

\subsection{Limitations}

This study and its conclusions are limited by several factors. Primarily, the method for network construction can be modified to the researcher's preference. Specifically, the threshold used to connected states within the Political Adjacency, Mitigation Similarity and Incidence Similarity networks were chosen in an attempt to capture the significant relationships. However, a more dense network can be created by lowering the threshold and and sparser network can be created by raising it. While we do not expect this to drastically impact the results of this paper, it nevertheless will likely cause the results to deviate slightly.

Further, other networks might be created, such as complete networks for the Political Adjacency, Mitigation Similarity, and Incidence Similarity with weighted edges. This would have the effect of capturing the ``distance'' between two states rather than making their relationships binary. We expect similar results to those presented in the paper; however, in our view, this would not highlight the importance of the underlying graph structure as much as the analysis presented above.

Finally, a county-level model could be considered instead of the state-level one. As many counties and municipalities enforced their own restrictions and have political affiliations that do not necessarily align with their states, a county-level analysis might uncover a more nuanced relationship between the political parties and how urban and rural counties approached mitigation differently.

\subsection{Conclusions and Future Work}

This study illustrates the strong relationship between political similarity and mitigation policy similarity. This can be seen through both the analysis of the Pearson Correlation between state-level politics and their time-dependent mitigation policies and through the network-based analysis. The former highlights the general relationship between politics and COVID-19 mitigation while the later emphasizes the underlying relationships between the states. Further, the network-based analysis allows us to consider both geographic relationships and incidence similarity and compare how these influence the mitigation policy when compared to the political relationships. We see that both similar disease-intensity and geographic distance between states state showed little to no correlation with similar mitigation policies after May of 2020 highlighting how politics became the driving factor behind controlling COVID-19 rather than the severity of the disease.

As discussed above, further work should consider other types of network constructions. These should be on both the county-level and using different criteria for connecting states together. Qualitative analysis would also offer a more nuanced perspective to the analysis. Consideration of interactions between state-level government officials could present explanations as to why politics were significantly tied to mitigation policy between June 2020 and June 2021.

Further work should also consider how social events, such as holidays, and other collective behavioral phenomena could influence the adjustment of policy. Finally, with a method to quantify each state's response to the pandemic over time, it can be important to understand to what extent the changes in policy changes diffuse through the networks and what factors drive said diffusion. Ultimately, we see that political similarity, not geographic proximity or pandemic intensity, played the most important role in predicting similar mitigation policies between states at the height of the pandemic.

\bibliographystyle{ieeetr}
\bibliography{Pol}

\end{document}